\documentclass[runningheads]{llncs}
\usepackage[T1]{fontenc}
\usepackage{graphicx}
\usepackage{color}

\usepackage{booktabs}
\usepackage{xcolor}
\usepackage{enumitem}
\usepackage[most]{tcolorbox}
\usepackage{array}
\usepackage{multirow}
\usepackage{multicol}
\usepackage{hyperref}
\usepackage{subcaption}

\begin{document}

\title{Echoes in the Sky: Computational Thematic Analysis of Online Public Discourse on Bluesky Across Trump's Reelection}
\titlerunning{Echoes in the Sky} 

\author{Qile Wang*\inst{1}\orcidID{0000-0003-0308-6033} \and
Ali Salloum*\inst{2}\orcidID{0000-0002-2381-6876} \and
Carolina Coimbra Vieira*\inst{3}\orcidID{0000-0003-3156-4151} \and
Benjamin E. Bagozzi\inst{1}\orcidID{0000-0002-6233-6453} \and
Mikko Kivelä \inst{2}\orcidID{0000-0003-2049-1954} \and
Kenneth E. Barner\inst{1}\orcidID{0000-0002-0936-7840} \and
Matthew Louis Mauriello\inst{1}\orcidID{0000-0001-5359-6520}} 
\authorrunning{Wang et al.}
%
\institute{University of Delaware, Newark, USA \\
\email{\{kylewang,bagozzib,barner,mlm\}@udel.edu} \and
Aalto University, Espoo, Finland \\
\email{\{ali.salloum,mikko.kivela\}@aalto.fi} \and
Max Planck Institute for Demographic Research, Rostock, Germany\\
\email{coimbravieira@demogr.mpg.de} 
}

\begingroup
\renewcommand\thefootnote{}\footnotetext{* These authors contributed equally.}
\endgroup

\maketitle              

\begin{abstract}
As political disruption intensifies online discourse, Bluesky has become an important platform for political discussion and public reaction. In this study, we examine large-scale discourse on Bluesky related to U.S. policy developments associated with the Trump administration. Using the historical retrieval API, we collected all available posts matching Trump and related keywords from 2019 to 2026, yielding 38.5 million posts. We leverage a large language model (LLM)-assisted clustering pipeline, combined with human validation, to identify 14 interpretable thematic domains in English-language posts and 19 thematic categories across 258 executive orders (EOs) signed between January 20, 2025, and May 1, 2026. Our findings identify several dominant themes in Bluesky discourse, including executive governance, political identity, and national security, as well as recurring themes in EOs, including executive task forces, border enforcement, and foreign policy. We also find substantial variation in the persistence and volatility of issue attention, accompanied by an increasing proportion of negative sentiment over time. The dataset and resources are publicly available at \url{https://github.com/Sensify-Lab/Echoes-in-the-Sky}


\end{abstract}

\section{Introduction}
Social media discourse increasingly reflects and shapes broader offline social and political dynamics \cite{conover2011political}. During periods of political disruption, online platforms often intensify and accelerate the circulation of public discourse \cite{jost2018social}. The reelection of President Donald Trump and the subsequent wave of U.S. policy shifts, including Executive Orders (EOs) 14162 (``Putting America First''), 14168 (``Gender Ideology''), 14257 (``Tariff''), and others, reignited polarized debate across offline and online spaces. While platforms such as X/Twitter have long served as central infrastructures for studying political communication and online behavior \cite{barbera2015tweeting}, recent API access restrictions and platform governance changes have constrained large-scale data collection through APIs. Consequently, alternative platforms such as Bluesky have emerged as promising environments for studying contemporary online political discourse and evolving digital publics \cite{salloum2025politics}. In this study, we investigate the online public discourse on Bluesky following Trump’s 2024 reelection, focusing on answering two main research questions: \textbf{(RQ1)} \textit{What are the main topics related to Trump discussed on Bluesky, and what sentiments are associated with them?} And, \textbf{(RQ2)} \textit{how do topics discussed on Bluesky evolve over time, and do major shifts coincide with Trump's EOs?}

\begin{figure}[!tp]
\centering
\includegraphics[width=0.99\linewidth]{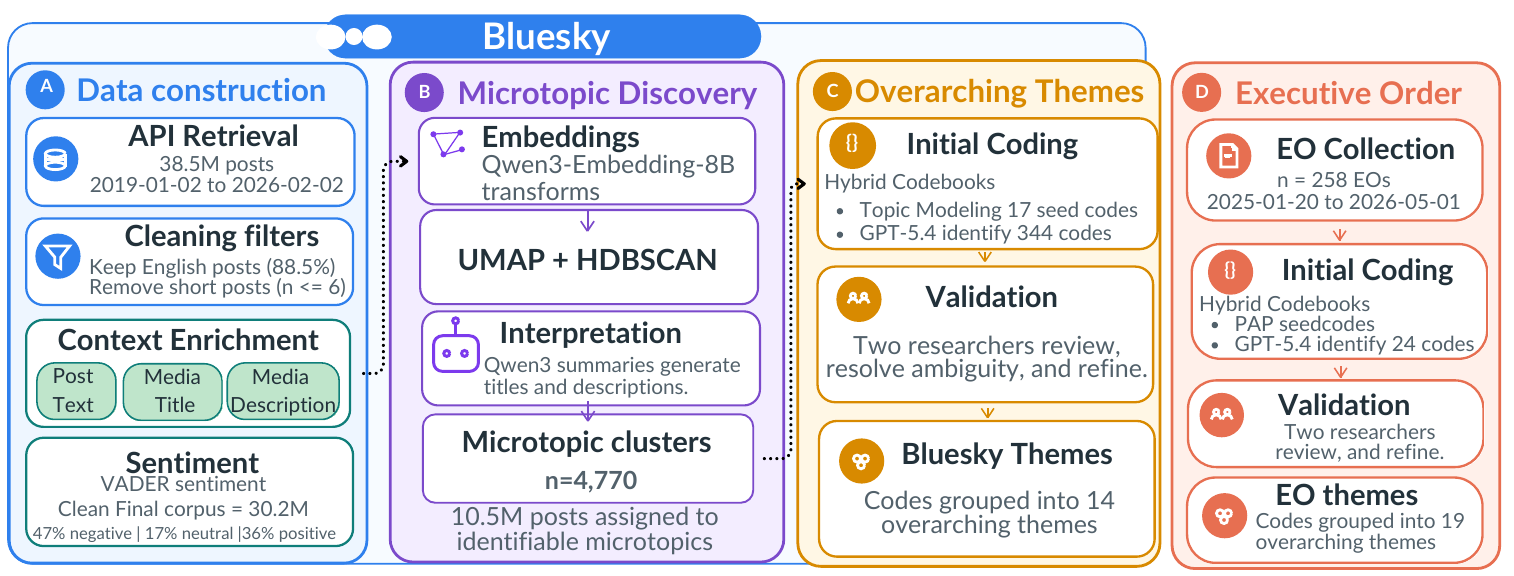}
\caption{Overview of the analysis pipeline, from Bluesky data retrieval and microtopic modeling to EO coding and theme alignment.}
\label{fig:pipeline}
\end{figure}

To answer these questions, we collected 38.5 million Bluesky posts related to President Donald Trump, spanning January 2019 to February 2026 (see overall method in Figure \ref{fig:pipeline}). We then characterize the thematic structure of Trump-related Bluesky posts and compare it with the thematic distribution of Trump-signed EOs. Our approach combines microtopic exploration, multi-LLM annotation, and human-in-the-loop thematic analysis. Next, we examine how attention to and sentiment toward overarching themes vary over time on Bluesky, focusing on whether some themes remain persistent while others appear episodically during periods of heightened political activity. Finally, we apply structural break detection to identify significant changes in discussion dynamics and evaluate whether these temporal shifts are associated with the timing of EOs. This work contributes by providing: (i) one of the largest publicly available research-oriented collections of Trump-related Bluesky posts; (ii) a scalable qualitative computational thematic analysis framework incorporating multi-LLM annotation and human validation; and (iii) an EO-linked annotation framework for studying the temporal dynamics of policy-driven online political discourse.

\section{Related Work}
Early studies examining Bluesky suggest that the platform exhibits strong political clustering and a predominantly left-leaning user base \cite{quelle2025bluesky,salloum2025politics}. However, few studies examine Bluesky discourse dynamics across sentiment, topic, and temporal dimensions. At the same time, rapid advances in LLMs have enabled new approaches to large-scale qualitative analysis \cite{wang2025leveraging,wang2026wisdom}. Prior research on Trump-related discussions across social media platforms suggests that online discourse can capture politically meaningful opinion trends and event-driven shifts in public attention. Twitter opinion trends during the 2016 U.S. presidential election closely tracked national polling aggregates \cite{bovet2018validation}, while Telegram discussions during the 2024 U.S. election responded to major political events \cite{paoletti2025tracing}. Building on these studies, we examine Trump-related discourse on Bluesky using sentiment analysis, topic modeling, LLM-assisted thematic analysis, and temporal analysis.

\section{Data and Methods}

\subsection{Bluesky}

\subsubsection{Data Collection}
We queried the API endpoint (\texttt{app.bsky.feed.searchPosts}) using a keyword lexicon designed to capture posts related to Donald Trump.\footnote{Keywords: ``trump'', ``maga'', ``trumpism'', ``potus45'', ``america first'', ``keep america great'', ``Trumpocalypse'', ``Republican'', ``Drumpf'', ``45th President''} 
To overcome the API limit of 100 items per request ($\approx$100,000 posts), we implemented a dynamic time-windowing strategy. Whenever a query reached this limit, the collection window was automatically reduced and the query rerun, with a minimum temporal resolution of one second. Each selected keyword was queried independently. After removing duplicate records, the final corpus comprised \textbf{38.5 million} posts. The collection period spans January 2, 2019, to February 2, 2026, capturing discourse pre- and post-Trump reelection. 
Accordingly, the dataset includes 241,389 posts in 2023, followed by rapid growth to 4.6 million posts in 2024 and 30.9 million posts in 2025, indicating substantial platform expansion (see Figure 1 in the Supplementary Material (SM)\footnote{\url{https://github.com/Sensify-Lab/Echoes-in-the-Sky/blob/main/SM.pdf}}).
We captured 1,364,569 unique authors, with each user contributing 28.2 posts ($SD=256.1$), representing roughly 3\% of global Bluesky users (see Table 1 in the SM).

\subsubsection{Data Preprocessing and Sentiment Analysis}
We constructed the analytic corpus using a rigorous multi-stage data-cleaning pipeline, as shown in Figure \ref{fig:pipeline}. First, we restricted our analysis to English-language posts (88.45\%, 34M posts). We cleaned posts by removing URLs, hashtags, and mentions while preserving emojis, given their linguistic value \cite{kejriwal2021empirical}. We then used the cleaned post text to estimate user sentiment with VADER.\footnote{\url{https://pypi.org/project/vaderSentiment/}}
Following \cite{gkikas2022text,baik2025analyzing}, we removed posts with fewer than six words and, when available, concatenated media titles and descriptions to provide contextual information for posts with limited text. This resulted in a cleaned English-language corpus of approximately \textbf{30.2 million} posts.

\subsubsection{Post-Level Microtopic Analysis}
\label{microtopic}

To identify fine-grained thematic structures within the corpus, we conducted a microtopic analysis at the individual post level, building on the prior framework \cite{salloum2025politics}. We partitioned the dataset into four subsets to preserve temporal patterns while enabling efficient processing and identification of topic emergence and decline.
The microtopic pipeline used ``Qwen3-Embedding-8B'' for semantic embeddings, UMAP for dimensionality reduction, and HDBSCAN for clustering to identify semantically coherent post clusters. Implementation details are available in the project repository.\footnote{\url{https://github.com/alesalloum/microtopics-analysis}}
To evaluate clustering robustness, we conducted hyperparameter exploration. Table 2 in the SM reports the optimal hyperparameters for microtopic clustering across temporal partitions.
To interpret clusters as microtopics, we analyzed their representative posts using frequent keywords and LLM-generated summaries. In the final stage, we extracted representative keywords with CountVectorizer and sampled up to 100 posts from each discovered cluster for semantic interpretation. These samples were processed by a vLLM-hosted language model (``30B-A3B-Instruct-2507’’\footnote{\url{https://huggingface.co/Qwen/Qwen3-30B-A3B-Instruct-2507}}), which generated concise titles and thematic descriptions for each cluster. The resulting microtopics were then reviewed and refined through human validation to support higher-level overarching thematic analysis.

\subsubsection{Overarching Theme Generation}
While microtopics reveal localized discourse patterns, higher-level themes are needed to interpret broader political narratives. After identifying microtopics, we developed a human-AI thematic analysis framework to group them into overarching themes, following thematic analysis principles \cite{braun2006using} and prior work on LLM-assisted qualitative synthesis with human validation \cite{wang2025lata}. The framework involved two steps: initial coding and theme development. In the initial coding stage, GPT-5.4 mini generated one or more candidate codes for each microtopic based on its title and description. To guide this process, we used a preliminary codebook of 17 seed codes derived from an independent topic modeling analysis of microtopic summaries (see Figure 2 in the SM), while allowing the model to generate new codes when existing codes were insufficient. Two researchers iteratively reviewed the candidate codes, resolved ambiguous cases, and refined code definitions. 

In the theme development stage, semantically related codes were grouped into approximately 10 to 25 candidate overarching themes. Each microtopic was then assigned to a single dominant theme to support macro-level interpretation and downstream temporal analysis. The researchers inspected the resulting theme structure and merged, renamed, or revised themes to improve conceptual clarity, distinctiveness, and interpretability. This iterative process combined model-generated semantic grouping with human interpretive judgment to produce the final set of overarching themes.

\subsection{Executive Orders (EOs)}
To study the association between themes discussed online on Bluesky and EOs, we conducted a separate thematic analysis of EOs signed during Donald Trump's second term. EO data were collected from the Federal Register Presidential Documents page\footnote{\url{https://www.federalregister.gov/presidential-documents/executive-orders}} and include 258 EOs signed between January 20, 2025, and May 1, 2026. Following the Bluesky theme-generation procedure, we used a hybrid coding strategy. The U.S. Policy Agendas Project (PAP) annotation scheme served as the initial codebook\footnote{\url{https://www.comparativeagendas.net/project/us/research/92}} and two researchers then reviewed and revised the annotations to ensure that the final EO themes accurately captured the substantive content of each order.

\subsection{Temporal Analysis}
To detect significant changes in thematic activity and the timing of shifts in user attention, we applied structural break analysis. We constructed daily post-volume time series for each Bluesky theme beginning on January 20, 2025, corresponding to the start of Trump’s second presidential term. For each time series $y_t$, we estimated breakpoints using the \texttt{strucchange} package in R. The procedure uses dynamic programming to identify breakpoints that minimize the residual sum of squares across segments: \( RSS = \sum_{j=1}^{K+1} \sum_{t \in S_j} (y_t - \bar{y}_j)^2 \) where $S_j$ denotes segment $j$, $\bar{y}_j$ is the mean daily post volume within segment $j$, and $K$ is the number of estimated breakpoints. The optimal number of breakpoints was determined by minimizing the Bayesian Information Criterion (BIC). Approximate confidence intervals for breakpoint locations were computed 
and reported when statistically meaningful interval estimates were available.

\section{Results \& Discussion}

\subsection{Sentiment Analysis}
We examine sentiment dynamics to assess whether Trump-related discussions are predominantly positive, negative, or neutral, and how these patterns change over time. We computed VADER sentiment scores using only the original user-authored post content,\footnote{Including media metadata changed sentiment classifications of only 4\% of posts.} excluding text from externally linked media to ensure that the sentiment scores reflect users' own expressions. Using this approach, 47.0\% of posts were classified as negative, 17.3\% as neutral, and 35.7\% as positive.

\begin{figure}[tb]
    \centering
    \includegraphics[width=0.45\textwidth]{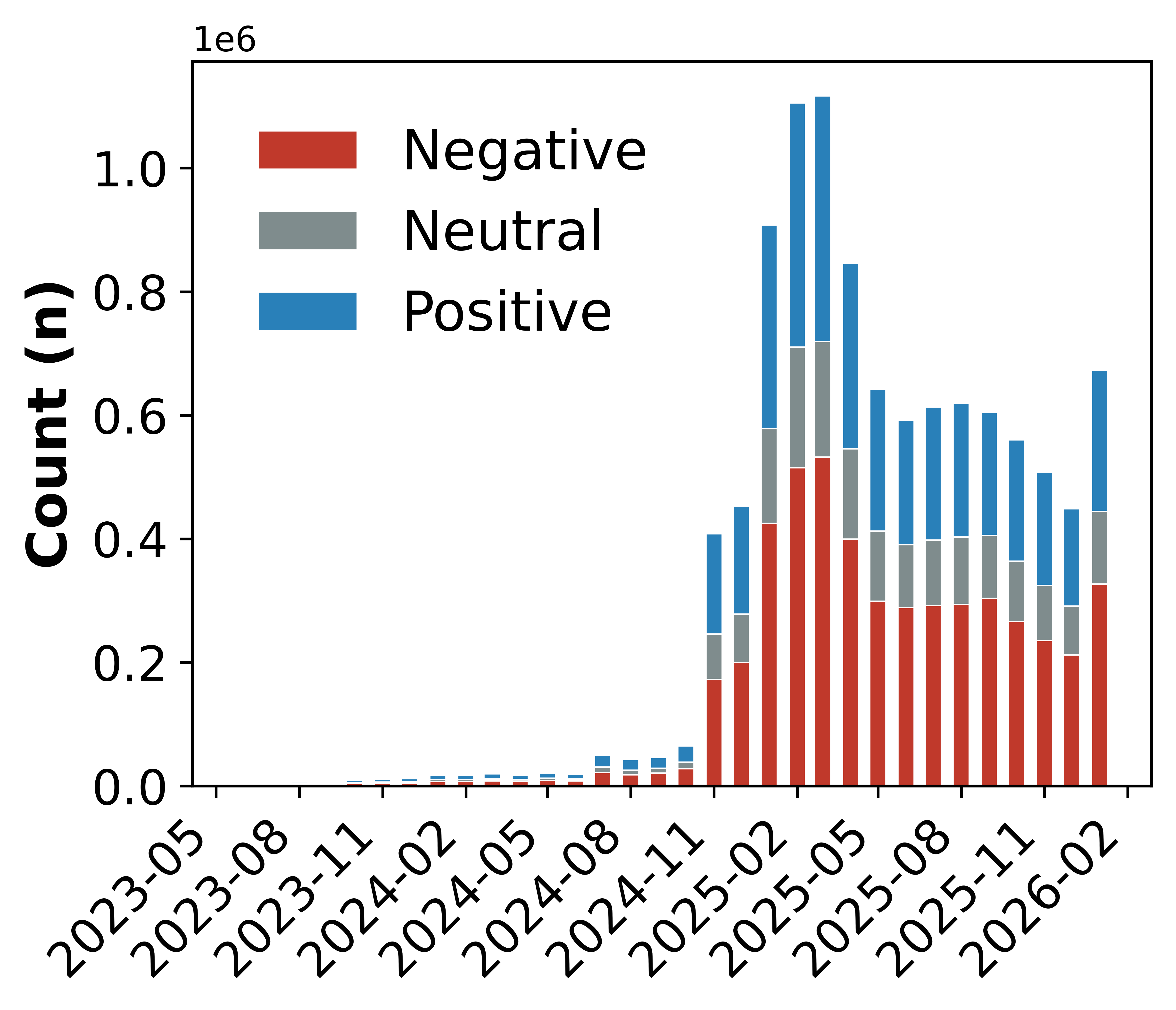}
    ~
    \includegraphics[width=0.45\textwidth]{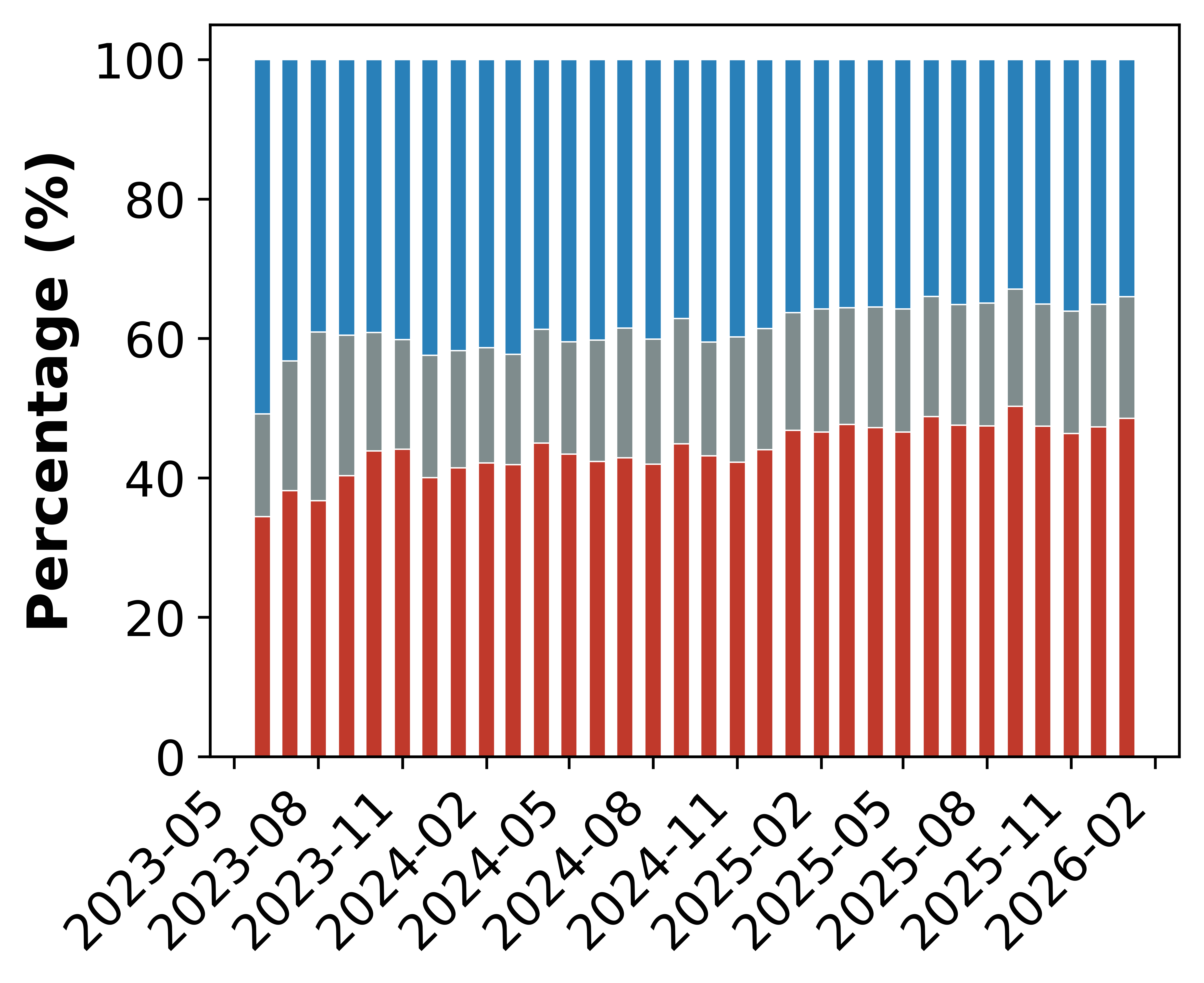}
    \caption{Monthly Sentiment Frequency. Monthly post counts by sentiment category (left) and sentiment shares as percentages of total posts (right).}
    \label{fig:monthly_sentiment_overview}
\end{figure}

Figure \ref{fig:monthly_sentiment_overview} shows the monthly distribution of posts by sentiment. Total Trump-related posting volume also increased sharply, with particularly large volumes in January and February 2025, coinciding with the start of Trump’s second presidential term. Overall, the figure suggests simultaneous growth in platform activity and the increasing dominance of negative sentiment in the monthly sentiment composition. Negative sentiment consistently represented the largest share of monthly Bluesky posts about Trump, while positive sentiment remained lower across most months. The gap between negative and positive sentiment increased over time, rising from 2\% in August 2024 to approximately 15\% by January 2026. Strongest negative sentiment gaps post-January 2025 appear in Public Health (45\%), Immigration (44\%), Public Lands (42\%), and National Security (34\%), reflecting Bluesky's left-leaning user composition and critical reactions to Trump administration policies (see Figures 3--17 in the SM).

\begin{table}[tb]
\centering
\caption{Examples of high-frequency microtopics identified across temporal partitions (A--D) and their corresponding overarching theme (B1--B14).}
\setlength{\tabcolsep}{3pt}
\renewcommand{\arraystretch}{0.92}
\resizebox{\textwidth}{!}{
\begin{tabular}{p{0.9cm} p{0.85cm} p{8cm} l p{1cm}}
\toprule
Part. & ID & Microtopic Summarization & Count & Theme \\
\midrule
A & 44 & Trump's Gaza Takeover Plan and Its Consequences & 184,591 & B3 \\
A & 809 & Trump's TikTok Paradox: From Ban to Savior & 47,579 & B5 \\
A & 35 & Project 2025: Trump's Hidden Agenda & 42,572 & B4 \\
A & 27 & Trump Blames DEI for Plane Crash Amid Safety Cuts & 42,206 & B1 \\
C & 1131 & Trump's Erosion of Disaster Response Systems & 39,904 & B9 \\
\bottomrule
\end{tabular}
}
\label{table:microtopic_examples_top20}
\end{table}

\subsection{Microtopic Discovery}
While sentiment captures the emotional tone of discourse, microtopic analysis identifies the substantive issues around which discussion is organized. Across the four temporal partitions, the optimized clustering pipeline identified 4,770 microtopic clusters. As a result, 10,456,083 posts were assigned to identifiable microtopics, while the remaining posts were treated as unclustered observations. Clustering quality metrics (silhouette scores 0.52--0.53, trustworthiness 0.95, assignment confidence 0.80--0.82) indicated stable, interpretable microtopics across temporal partitions.
Table \ref{table:microtopic_examples_top20} presents the highest-frequency microtopics identified across the temporal partitions. These microtopics capture a broad range of political discourse themes, including international conflict, electoral politics, executive governance, media controversies, partisan polarization, foreign policy, and disaster response. Several themes recur across partitions, particularly discussions of Gaza, Elon Musk, and the Republican Party. This recurrence suggests both continuity and evolution in public political attention over time. These fine-grained microtopics also provide the foundation for the subsequent macro-level thematic analysis, where related microtopics are merged into broader interpretive categories.

\begin{figure}[tb]
    \centering
    \includegraphics[width=1\textwidth]{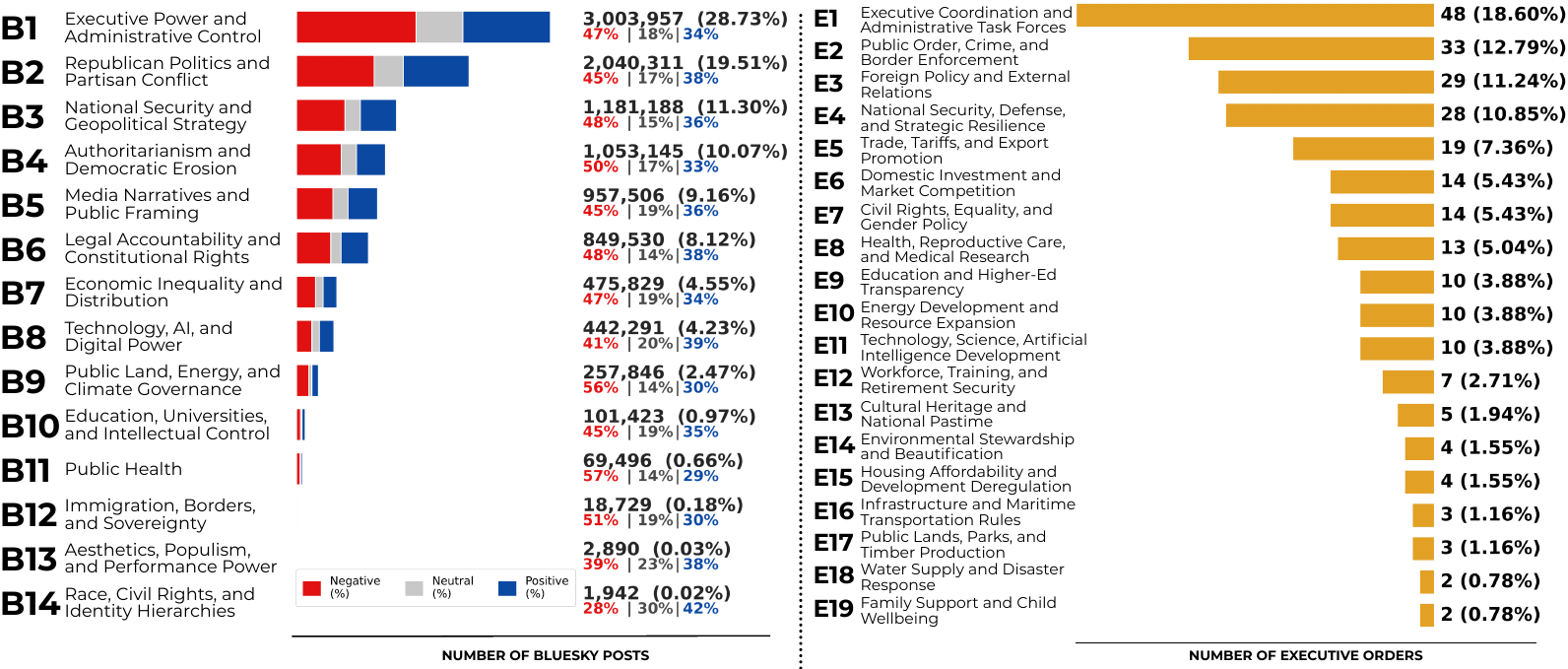}
    \caption{Distribution of overarching themes in \textbf{B}luesky posts (left) and \textbf{E}xecutive orders signed during Trump's second term (right). Percentages indicate the share of posts or EOs mapped to each theme, while colors in the left panel denote the sentiment composition of Bluesky posts within each theme.}
    \label{fig:theme_distribution}
\end{figure}

\subsection{EOs and Bluesky Overarching Themes Annotation}
\label{sec:theme_results}
Using our human-AI thematic analysis pipeline, we generated 344 initial codes from the microtopics and consolidated them into 14 overarching themes. The left panel of Figure \ref{fig:theme_distribution} presents the distribution of these Bluesky themes, along with their corresponding sentiment composition. Among the most prominent themes were Executive Power (B1) and Partisan Conflict (B2).
Similarly, we identified 19 thematic categories across 258 executive orders, as shown in the right panel of Figure \ref{fig:theme_distribution}. The largest category was Executive Coordination and Administrative Task Forces (18.60\%), followed by Public Order, Crime, and Border Enforcement (12.79\%), Foreign Policy and External Relations (11.24\%), and National Security, Defense, and Strategic Resilience (10.85\%). Trade, Tariffs, and Export Promotion was also prominent (7.36\%). These categories point to a strong emphasis on administrative coordination, enforcement, foreign policy, and security governance, closely mirroring the dominant themes in Bluesky discourse.

\begin{figure}[tb]
    \centering
    \includegraphics[width=0.95\textwidth]{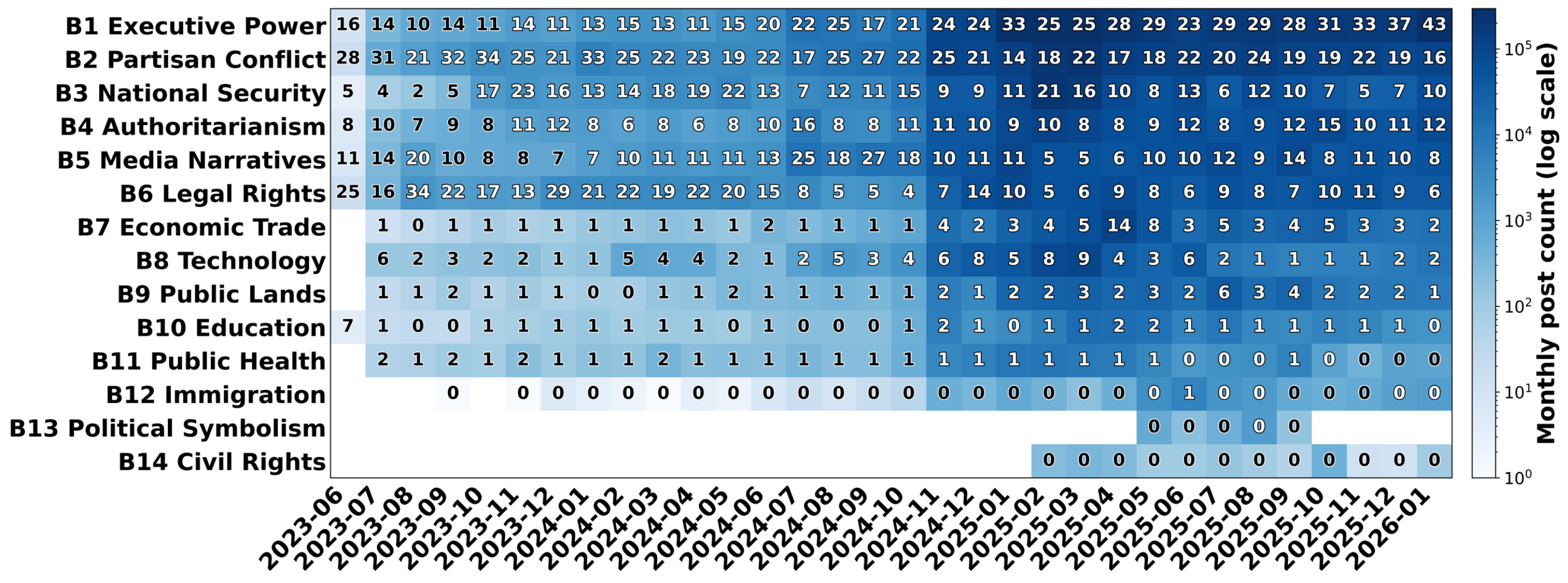}
    \caption{Temporal dynamics of Bluesky political discussion themes by month. The heatmap shows monthly post volumes for 14 themes on a logarithmic color scale, with cell values indicating the percentage of theme-assigned posts in that month. Y-axis labels use abbreviated theme names from Figure \ref{fig:theme_distribution} (left). }
    \label{fig:theme_monthly_count_heatmap}
\end{figure}

However, political attention is inherently dynamic. Some issues remain consistently salient, while others emerge briefly in response to specific events before rapidly declining. Examining how attention evolves across themes provides insight into how online publics allocate and reallocate attention during periods of political change. Figure \ref{fig:theme_monthly_count_heatmap} indicates the monthly evolution of Bluesky overarching themes. Discussion volume increased across nearly all themes after January 2025, coinciding with the start of Trump’s second term. Executive Power (B1) and Partisan Conflict (B2) remained dominant throughout the observation period. Several issue-specific themes also exhibited marked temporal variation. Economic Trade (B7) peaked (14\%) in April 2025, while Media Narratives (B5) and Authoritarianism (B4) were most pronounced in September and October 2025. These fluctuations suggest that thematic attention on Bluesky was closely responsive to contemporaneous political events and policy developments.

\subsection{Structural Break Analysis}
\begin{figure}[tb]
    \centering
    \includegraphics[width=0.95\textwidth]{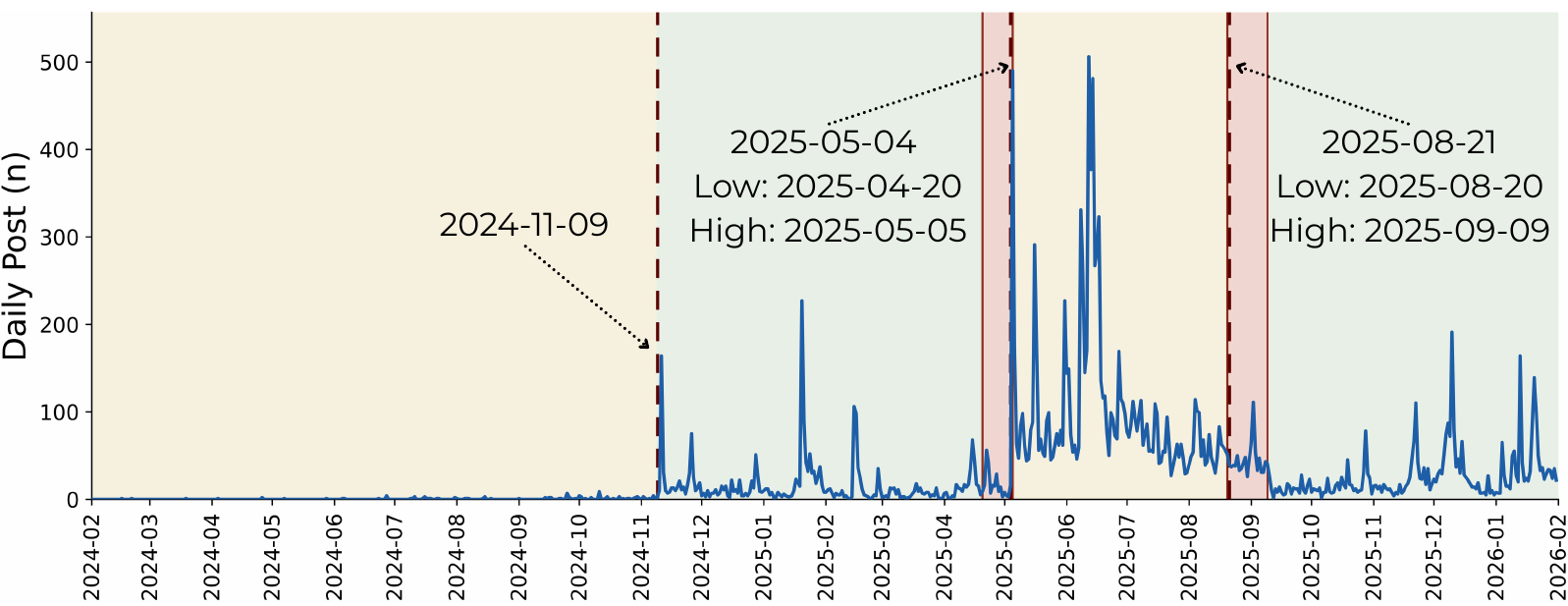}
    \caption{Structural breaks in the  Theme Immigration (B12). Dashed vertical lines mark estimated breakpoint dates, and shaded regions show 95\% confidence intervals where reliable estimates were available.}
    \label{fig:immigration}
\end{figure}

Visual inspection of temporal trends in Bluesky post volume reveals fluctuations in public attention, while structural break analysis identifies statistically significant changes in discussion dynamics. Figure \ref{fig:immigration} identifies several statistically significant structural breaks in immigration-related discourse. These breakpoints correspond to abrupt changes in posting activity and suggest transitions between distinct attention regimes. Several breaks occur near major border control executive actions, including EO 14287 (``Protecting American Communities From Criminal Aliens''), signed on April 28, 2025; EO 14321 (``Ending Crime and Disorder on America’s Streets''); and EO 14333 (``Declaring a Crime Emergency in the District of Columbia''), signed on August 11, 2025. These temporal alignments suggest that a combination of policy announcements, media coverage, and broader political debate may have shaped immigration-related discourse on Bluesky. Rather than evolving gradually, immigration discourse appears episodic, with attention shifting abruptly between relatively stable periods. Detected breakpoints cluster around discrete political events, such as executive actions and associated media coverage, suggesting these events coincide with abrupt changes in discussion volume. This episodic pattern is consistent with those observed across discourse themes analyzed (see Figures 18-26 in the SM).

\section{Conclusion \& Future Work}
In this study, we analyzed Trump-related Bluesky posts to examine how online political discourse evolves around major U.S. policy developments. Using an LLM-assisted thematic analysis pipeline with human validation, we identified discourse themes spanning executive governance, partisan conflict, immigration, international affairs, technology, media narratives, and democratic institutions. The findings show that Bluesky attention is both persistent and event-sensitive, with some themes remaining consistently salient and others rising episodically around major political or policy developments. Negative sentiment was also prominent throughout the corpus, especially after the beginning of Trump's second presidential term. Our analysis is limited to English-language Trump-related posts. The sentiment analysis is based on VADER and does not distinguish supportive from opposing sentiment. In addition to that, temporal associations should not be interpreted as causal. Future work should examine multilingual discourse, user-level dynamics, and stance.

\vspace{0.2cm}
\noindent\textbf{\ackname} 
This work was carried out while the first author was a PhD student and during a research visit to the Max Planck Institute for Demographic Research. We thank Emilio Zagheni, Tom Theile, Aliakbar Akbaritabar, Egor Kotov, and Abigail Tun Mendicuti for their conceptual guidance. 




%
%
%
\bibliographystyle{splncs04}
\bibliography{reference}

\end{document}